%% file: paper.tex
\documentclass[twocolumn,showpacs,aps,prl,superscriptaddress]{revtex4}
\usepackage{graphicx}
\usepackage{lineno}
\usepackage{dcolumn}
\usepackage{amsmath}
\usepackage{epsfig}
\usepackage{multirow}

\newcommand{\BaBarYear}       {06}
\newcommand{\BaBarNumber}     {058}
\newcommand{\SLACPubNumber} {12040}

\input pubboard/babarsym

\def\pipi     {\ensuremath{\pi\pi}}

\def\fpm {\ensuremath{f_{\pm}(\deltat)}}

\def\de {\ensuremath{\Delta E}}

\RequirePackage{xspace}

\def\figurebox#1#2#3{%
    \def\arg{#3}%
    \ifx\arg\empty
    {\hfill\vbox{\hsize#2\hrule\hbox to #2{\vrule\hfill\vbox to #1{\hsize#2\vfill}\vrule}\hrule}\hfill}%
    \else
    {\hfill\epsfbox{#3}\hfill}%
    \fi}

\long\def\inst#1{\par\nobreak\kern 4pt\nobreak
    {\it #1}\par\vskip 10pt plus 3pt minus 3pt}

\begin{document}

\begin{flushleft}
\babar-PUB-\BaBarYear/\BaBarNumber\\
SLAC-PUB-\SLACPubNumber
\end{flushleft}

\title{
{
\Large \bf \boldmath
Observation of $\Bu\to\Kzb\Kp$ and $\Bz\to\Kz\Kzb$
}}

\input pubboard/authors_jul2006

\begin{abstract}
We report observations of the $b\to d$ penguin-dominated decays 
$\Bu\to\Kzb\Kp$ and $\Bz\to\Kz\Kzb$ in $316\invfb$ of $e^+e^-$ collision data collected with the \babar\ detector. We measure the branching fractions 
$\BR(\Bu\to\Kzb\Kp) = (1.61 \pm 0.44 \pm 0.09)\times 10^{-6}$
and $\BR(\Bz\to\Kz\Kzb) = (1.08 \pm 0.28 \pm 0.11)\times 10^{-6}$, 
and the \CP-violating charge asymmetry 
${\cal A}_{\CP}\,(\Kzb\Kp) = 0.10\pm 0.26\pm 0.03$.  Using a 
vertexing technique previously employed in several analyses of all-neutral final states 
containing kaons, we report the first measurement of 
time-dependent \CP-violating asymmetries in $\Bz\to\KS\KS$, obtaining
$S = -1.28^{+0.80\;+0.11}_{-0.73\;-0.16}$ and
$C = -0.40\pm 0.41\pm 0.06$.  We also 
report improved measurements of the branching fraction 
$\BR(\Bu\to\Kz\pip) = (23.9\pm 1.1\pm 1.0)\times 10^{-6}$ and \CP-violating 
charge asymmetry ${\cal A}_{\CP}\,(\Kz\pip) = -0.029\pm 0.039\pm 0.010$.
\end{abstract}

\pacs{
13.25.Hw,
11.30.Er, 
12.15.Hh 
}

\maketitle

The decays $\Bu\to\Kzb\Kp$ and $\Bz\to\Kz\Kzb$ are 
expected to be dominated by the flavor-changing neutral-current process 
$b\to d\bar{s}s$, which is highly suppressed in the standard model and
potentially sensitive to the presence of new particles in a way analogous
to $b\to s\bar{s}s$ decays such as $B\to\phi K$~\cite{LondonQuinn,phiK}.
Assuming top-quark dominance in the virtual loop mediating the $b\to d$
transition~\cite{Fleischer94}, the charge asymmetry in $\Bu\to\Kzb\Kp$
and the time-dependent \CP-violating asymmetry parameters in $\Bz\to\KS\KS$
are expected to vanish, while contributions from lighter quarks or 
supersymmetric particles could induce observable asymmetries~\cite{Giri}.
It has been noted~\cite{FleischerRecksiegel} that the branching fraction and
\CP\ asymmetries in $\Bz\to\Kz\Kzb$ are related in a nearly model-independent way,
providing a sensitive test of the standard model description of \CP\ violation.

In this Letter, we report observations of $\Bu\to\Kzb\Kp$ and 
$\Bz\to\Kz\Kzb$ using a data sample approximately $50\%$ larger than the one
used in our previous search~\cite{KsX05}.  (The 
use of charge-conjugate modes is implied throughout this paper unless 
otherwise stated.)  In addition to establishing 
these decay modes, we present measurements of the time-dependent 
\CP-violating asymmetries in $\Bz\to\Kz\Kzb$ for the first time.  
We also report updated measurements of the branching fraction and 
charge asymmetry in the $SU(3)$-related decay $\Bu\to\Kz\pip$.

The \CP\ asymmetry in $\Bz\to\Kz\Kzb$ (observed in the $\KS\KS$ final state)
is determined from the difference in the time-dependent decay rates for
$\Bz$ and $\Bzb$.  In the process $\epem\to\Y4S\to\Bz\Bzb$, the decay rate
$f_+\,(f_-)$ is given by~\cite{tdcpv}
\begin{eqnarray}
\fpm = \frac{e^{-\left|\deltat\right|/\tau}}{4\tau} [1
& \pm & S\sin(\deltamd\deltat) \nonumber \\
& \mp & C\cos(\deltamd\deltat)]
\label{fplusminus}
\end{eqnarray}
when the second $B$ meson in the event (denoted $B_{\rm tag}$) is identified as
$\Bz\,(\Bzb)$.  Here $\Delta t$ is the time difference between the decays of
the signal and $B_{\rm tag}$ mesons, $\tau$ is the average $\Bz$ lifetime,
and $\Delta m_d$ is the $\Bz-\Bzb$ mixing frequency.
The amplitude $S$ describes \CP\ violation in the interference
between mixed and unmixed decays into the same final state, while $C$ describes
direct \CP\ violation in decay.

The data sample used in this analysis contains $(347.5\pm 3.8)\times 10^6$
$\Y4S\to\BB$ decays collected by the \babar\ detector~\cite{babar} at the
Stanford Linear Accelerator Center's (SLAC) \pep2\ asymmetric-energy $\epem$ collider.  The primary detector
elements used in this analysis are a charged-particle tracking system 
consisting of a five-layer silicon vertex tracker and a 40-layer drift 
chamber surrounded by a $1.5$-T solenoidal magnet, and a dedicated 
particle-identification system consisting of a detector of internally 
reflected Cherenkov light.

We identify two separate event samples corresponding to the decays 
$\Bu\to \KS h^+$ and $\Bz\to \KS\KS$, where $h^+$ is either a pion or a 
kaon.  Neutral kaons are reconstructed in the mode $\KS\to\pip\pim$ by 
combining pairs of oppositely charged tracks originating from a common decay 
point and satisfying selection requirements on their invariant mass and
proper decay time.  Candidate $h^+$ tracks are assigned the 
pion mass and are required to originate from the interaction region and to have 
a well-measured Cherenkov angle $(\theta_c$) consistent with either the
pion or kaon particle hypothesis.

For each $\Bz$ candidate, we require the absolute value of the difference $\de$ between its 
reconstructed energy in the center-of-mass (CM) frame and the beam energy $(\sqrt s/2)$ 
to be less than $100\mev$.  For $\Bu$ candidates, we require
$-115 < \de < 75\mev$, where the lower limit accounts for an average
shift in $\de$ of $-45\mev$ in the $\Kzb\Kp$ mode due to the assignment of 
the pion mass to the $\Kp$ candidate.  We also define a beam-energy substituted mass 
$\mes \equiv \sqrt{(s/2 + {\mathbf {p}}_i\cdot {\mathbf {p}}_B)^2/E_i^2 - 
{\mathbf {p}}_B^2}$, where the $B$-candidate momentum ${\mathbf {p}}_B$ 
and the four-momentum of the initial $\epem$ state 
$(E_i, {\mathbf {p}}_i)$ are calculated in the laboratory frame.  We require
$5.20 < \mes < 5.29\gevcc$ for $B$ candidates in both samples.  To suppress 
the dominant background arising from the process $\epem\to\qqbar\,~(q=u,d,s,c)$, 
we calculate the CM angle $\theta^*_S$ between the sphericity axis~\cite{sphericity}
of the $B$ candidate and the sphericity axis of the remaining charged and neutral 
particles in the event, and require $\left | \cos\theta^*_S \right | < 0.8$.

\renewcommand{\multirowsetup}{\centering}
\newlength{\LL}\settowidth{\LL}{$8047$}
\begin{table*}[!htb]
\begin{center}
\caption{Summary of results for the total detection efficiencies 
$\eps$, fitted signal yields $n$, signal-yield significances $s$ (including
systematic uncertainty), charge-averaged branching fractions $\BR$, and 
charge asymmetries ${\cal A}_{\CP}$ (including $90\%$ confidence intervals).  
The efficiencies include the branching fraction for
$\KS\to\pip\pim$ and the probability of $50\%$ 
for $\Kz\Kzb\to \KS\KS$.  Branching fractions are calculated 
assuming equal rates for $\upsbzbz$ and $\Bp\Bm$~\cite{BBratio}.}
\label{tab:summary}
\begin{ruledtabular}
\begin{tabular}{lcccccc}
Mode  & $\eps$ (\%) & $n$ & $s~(\sigma)$ & \BR~($10^{-6}$) & ${\cal A}_{\CP}$ &
${\cal A}_{\CP}\,(90\% {\rm CL})$\\
\hline \\[-3mm]
$\Bu\to\Kz\pip$  & $12.9 \pm 0.4$ & $1072\pm 46\,^{+32}_{-37}$  &
           & $23.9 \pm 1.1 \pm 1.0$ & $-0.029\pm 0.039\pm 0.010$ & $[-0.092,0.036]$\\[2mm]
$\Bu\to\Kzb\Kp$  & $12.6 \pm 0.4$ & $71\pm 19\pm 4$ &
            $5.3$ & $1.61\pm 0.44\pm 0.09$ & $0.10\pm 0.26\pm 0.03$ & $[-0.31,0.54]$ \\[2mm]
$\Bz\to\KzKzb$   & $8.5 \pm 0.3$ & $32\pm 8\pm 3$ &
            $7.3$ & $1.08\pm 0.28\pm 0.11$ & \\
\end{tabular}
\end{ruledtabular}
\end{center}
\end{table*}

After applying all of the above requirements, we find $2321$ $(30159)$
candidates in the $\Bz\,(\Bu)$ sample.  The total 
detection efficiencies are given in Table~\ref{tab:summary} and
include the branching fraction for $\KS\to\pip\pim$~\cite{PDG2004} 
and a probability of $50\%$ for $\Kz\Kzb\to \KS\KS$~\cite{KsKl}.
We use data and simulated Monte Carlo samples~\cite{geant} to verify that 
backgrounds from other $B$ decays are negligible.

A multivariate technique~\cite{Sin2betaPRD} is employed to determine the 
flavor of the $B_{\rm tag}$ meson in the $\Bz$ sample.  Separate neural 
networks are trained to identify primary leptons, kaons, low-momentum pions from 
$D^*$ decays, and high-momentum charged particles from \B\ decays.  Events 
are assigned to one of six mutually exclusive ``tagging'' categories. 
The quality of tagging is expressed in terms of the 
effective efficiency $Q = \sum_k \epsilon_k (1-2w_k)^2$, where $\epsilon_k$ 
and $w_k$ are the efficiencies and mistag probabilities, respectively, for events tagged 
in category $k$.  We measure the tagging performance in a data sample of 
fully reconstructed neutral $B$ decays ($B_{\rm flav}$) to 
$D^{(*)-}(\pip,\, \rho^+,\, a_1^+)$, where the flavor of the decaying $B$ meson is known, 
and find a total effective efficiency of $Q = (30.4\pm 0.3)\%$.

The time difference $\deltat \equiv \Delta z/\beta\gamma c$ is obtained 
from the known boost of the $\epem$ system ($\beta\gamma = 0.56$) and the 
measured distance $\Delta z$ along the beam ($z$) axis between the 
$\Bz\to\KS\KS$ and $B_{\rm tag}$ decay vertices.  The position of the 
$B_{\rm tag}$ vertex is determined from the remaining charged particles 
in  the event after removing the four tracks composing the signal candidate.
Despite the relatively long lifetime of the $\KS$ mesons, the $z$ position
of the $B$-candidate decay point is obtained reliably by exploiting the precise
knowledge of the interaction point using the technique described in
Ref.~\cite{KsVertexing}.  We compute $\deltat$ and its error
from a combined fit to the $\Y4S\to\Bz\Bzb$ decay, including the constraint
from the known average lifetime of the $\Bz$ meson.  Approximately $82\%$ of signal events contain a $\KS$ reconstructed from pions that each have at least two hits in the silicon vertex tracker, providing sufficiently small $\deltat$ uncertainty ($0.9\,{\rm ps}$) to perform the measurement.  We require $\left|\deltat\right|<20\ps$ and 
$\sigma_{\deltat} < 2.5\ps$, where $\sigma_{\deltat}$ is the uncertainty 
on $\deltat$ determined separately for each event.  The resolution function 
for signal candidates is a sum of three gaussian distributions with 
parameters determined from the $B_{\rm flav}$ sample~\cite{Sin2betaPRD}.  
The background $\deltat$ distribution has the same functional form as the 
signal resolution function, with parameters determined directly 
from data.

To obtain the yields and \CP\-violating asymmetry parameters in each sample, we 
apply separate unbinned maximum-likelihood fits incorporating 
discriminating variables that account for differences between $\BB$ and 
$\qqbar$ events.  In addition to the kinematic variables $\mes$ and $\de$, 
we include a Fisher discriminant ${\cal F}$~\cite{pipi2002} defined as 
an optimized linear combination of the event-shape variables 
$\sum_i p_i^*$ and $\sum_i p_i^*\cos^2\theta_i^*$, where $p_i^*$ is 
the CM momentum of particle $i$, $\theta_i^*$ is the CM angle between 
the momentum of particle $i$ and the $B$-candidate thrust axis, and the 
sum is over all particles in the event excluding the $B$ daughters.  For
the fit to the $\Bu$ sample we include the Cherenkov angle measurement
to separate $\KS\pip$ and $\KS\Kp$ decays.  For the $\Bz$ sample we include 
$\deltat$ to determine the $\CP$-violating asymmetry parameters $S$ and $C$ simultaneously with the signal yield.

The likelihood function to be maximized is defined as
${\cal L} = \exp{\left (-\sum_{i}n_i \right )}
\prod_{j=1}^N\left[\sum_i n_i{\cal P}_i\right]$,
where $n_i$ and ${\cal P}_i$ are the yield and probability density
function (PDF) for each component $i$ in the fit, and $N$ is the total
number of events in the sample.  For the $\Bz$ sample there
are two components (signal and background), and the total PDF is
calculated as the product of the individual PDFs for $\mes$, $\de$,
${\cal F}$, and $\deltat$.  The signal $\deltat$ PDF is derived from 
Eq.~\ref{fplusminus}, modified to take into account the mistag probability 
and convolved with the resolution function.
We combine $\Bu$ 
and $\Bub$ candidates in a single fit and include the PDF for 
$\theta_c$ to determine separate yields and charge asymmetries for 
the two signal components, $\KS\pi$ and $\KS K$, and two corresponding 
background components.  For both signal and background, the 
$\KS h^{\pm}$ yields are parameterized as 
$n_{\pm} = n(1 \mp {\cal A}_{\CP})/2$; we fit directly for the total 
yield $n$ and the charge asymmetry ${\cal A}_{\CP}$.  We have found correlations among the PDF variables in the fit to be negligible in both the $\Bz$ and $\Bp$ samples.

The parameterizations of the PDFs are determined from data wherever
possible.  In both samples, we exploit the large sideband regions in 
$\mes$ and $\de$ to determine all background PDF parameters simultaneously 
with the yields and \CP\ asymmetries in the fits.  For the $\Bu$ sample, 
the large signal $\KS\pip$ component allows for an accurate determination 
of the peak positions for $\mes$ and $\de$, as well as the parameters 
describing the shape of the PDF for ${\cal F}$.  The remaining shape 
parameters describing $\mes$ and $\de$ are determined from simulated 
Monte Carlo samples and are fixed in the fit.  We use the $\KS\pip$ 
parameters to describe signal $\KS\Kp$ PDFs in $\mes$, $\de$, and 
${\cal F}$, taking into account the known shift in the mean of $\de$ 
due to the pion-mass hypothesis.  For both signal and background, the 
$\theta_c$ PDFs are obtained from a sample of 
$D^{*+}\to \Dz\pip\,(\Dz\to\Km\pip)$ decays 
reconstructed in data, as described in Ref.~\cite{KpiBaBar}.
For the $\Bz$ sample, all shape parameters describing the $\mes$, $\de$, and 
${\cal F}$ signal PDFs are fixed to the values determined from Monte Carlo simulation except
the peak position for $\de$, which is derived from the 
results of the fit to the $\Bu$ sample.

Several cross-checks were performed to validate the fitting 
technique before data in the signal region were examined.  
We checked for biases by performing pseudo-experiments where simulated 
Monte Carlo signal events were mixed with background events generated 
directly from the PDFs according to the expected yields in the data.
The resulting small biases on the yields include effects of incorrect particle
identification and are accounted for in the systematic uncertainties.

The fit results supersede our previous measurements of these
quantities and are summarized in Table~\ref{tab:summary}.
The signal yields for $\Bu\to\KS\Kp$ and $\Bz\to\KS\KS$ correspond 
to significances of $5.3\sigma$ and $7.3\sigma$ (including
systematic uncertainties), respectively, and are consistent with our 
previous measurements~\cite{KsX05}, as well as with recent results by
the Belle Collaboration~\cite{BelleKK}.  The significances are computed by taking the square root of the change in $2$ln${\cal L}$ when the appropriate yield is fixed to zero. The fit to the $\Bz$ sample yields $S = -1.28^{+0.80\;+0.11}_{-0.73\;-0.16}$ and $C = -0.40\pm 0.41\pm 0.06$, where the first errors are statistical and the second are systematic.  The linear correlation coefficient between $S$ and $C$ is $-32\%$.

\begin{figure}[!tbp]
\begin{center}
\includegraphics[width=0.5\linewidth]{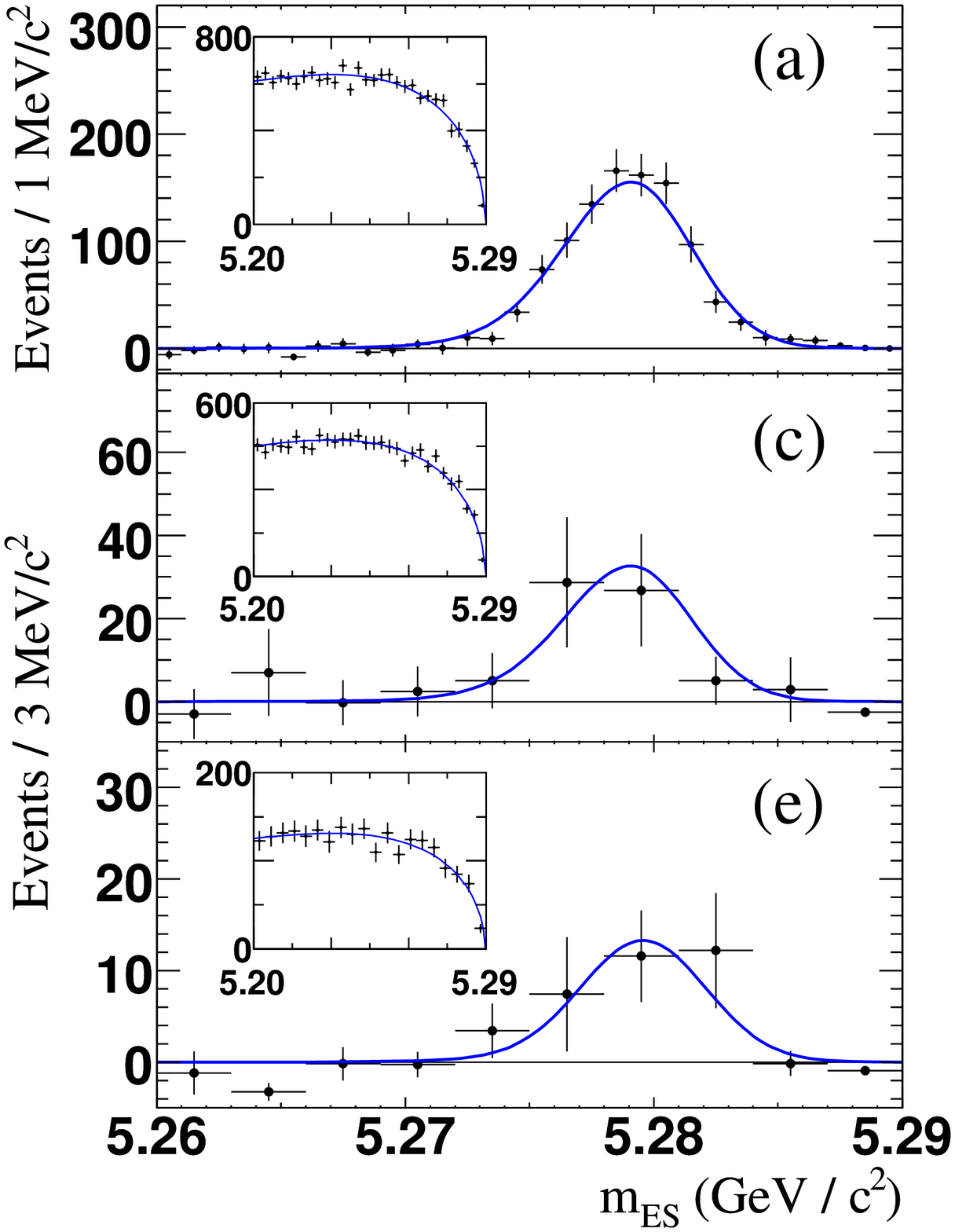}\includegraphics[width=0.5\linewidth]{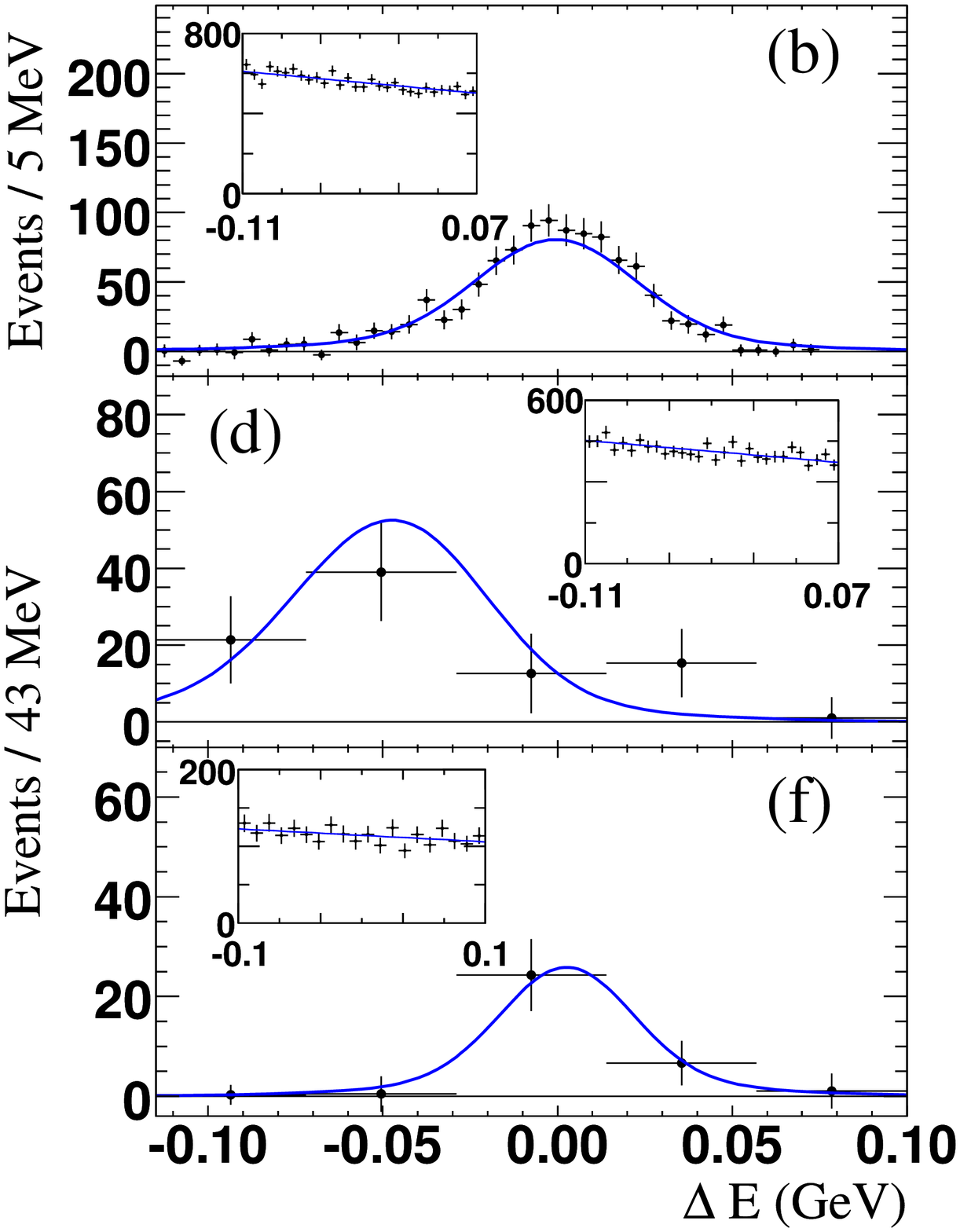}
\caption{Distributions of (left) $\mes$ and (right) $\de$ for signal (main plot) and background (inset)
(a),(b) $\KS\pi ^+$, (c),(d) $\KS K^+$, and (e),(f) $\KS\KS$ candidates (points with error bars) in 
data obtained with the weighting technique described in the text.  The solid curves represent 
the assumed shapes used in the fits.}
\label{fig:sPlots}
\end{center}
\end{figure}

\begin{figure}[!tbp]
\begin{center}
\includegraphics[height=5.8cm,width=0.50\linewidth]{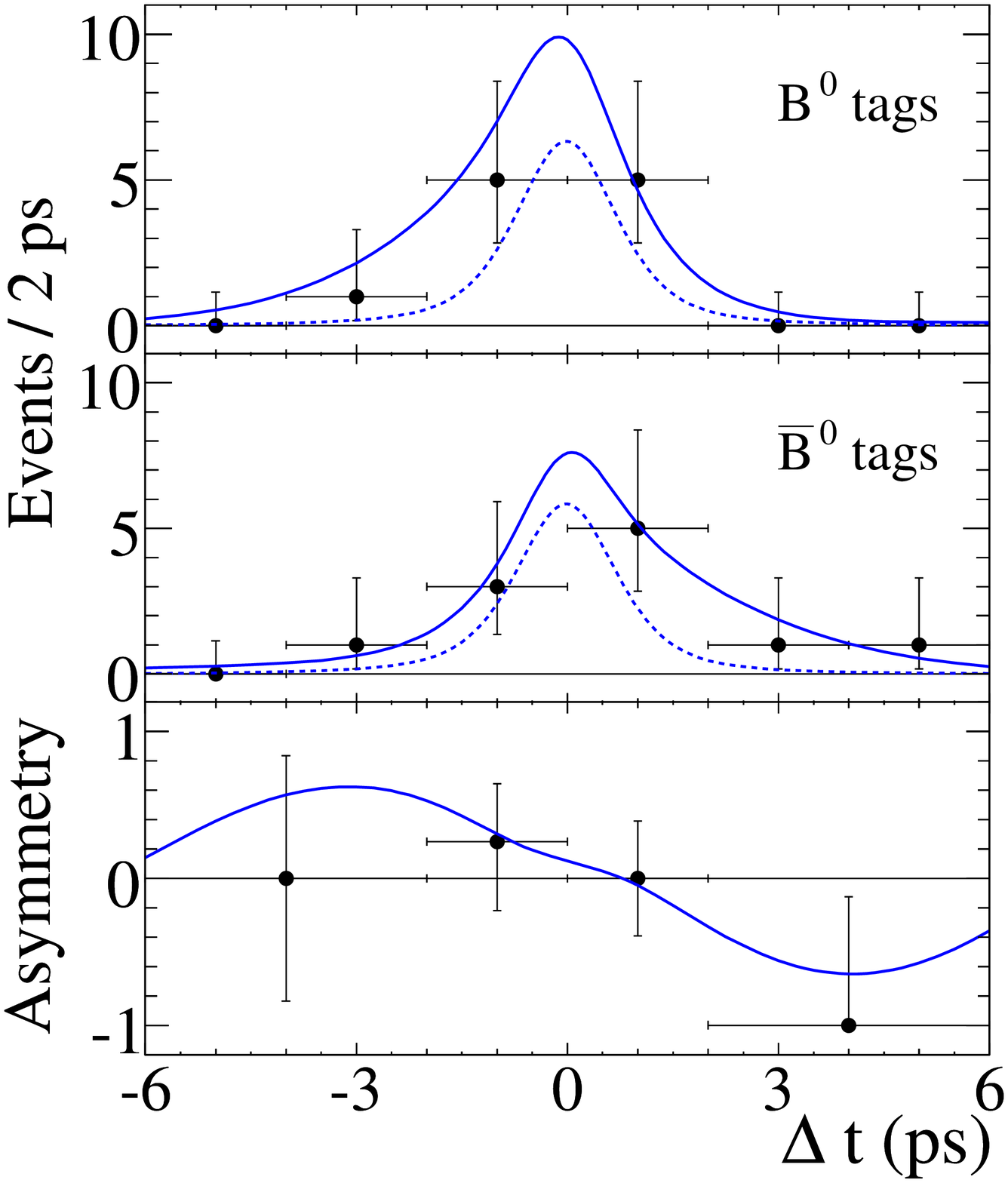}\hfill\includegraphics[width=0.50\linewidth]{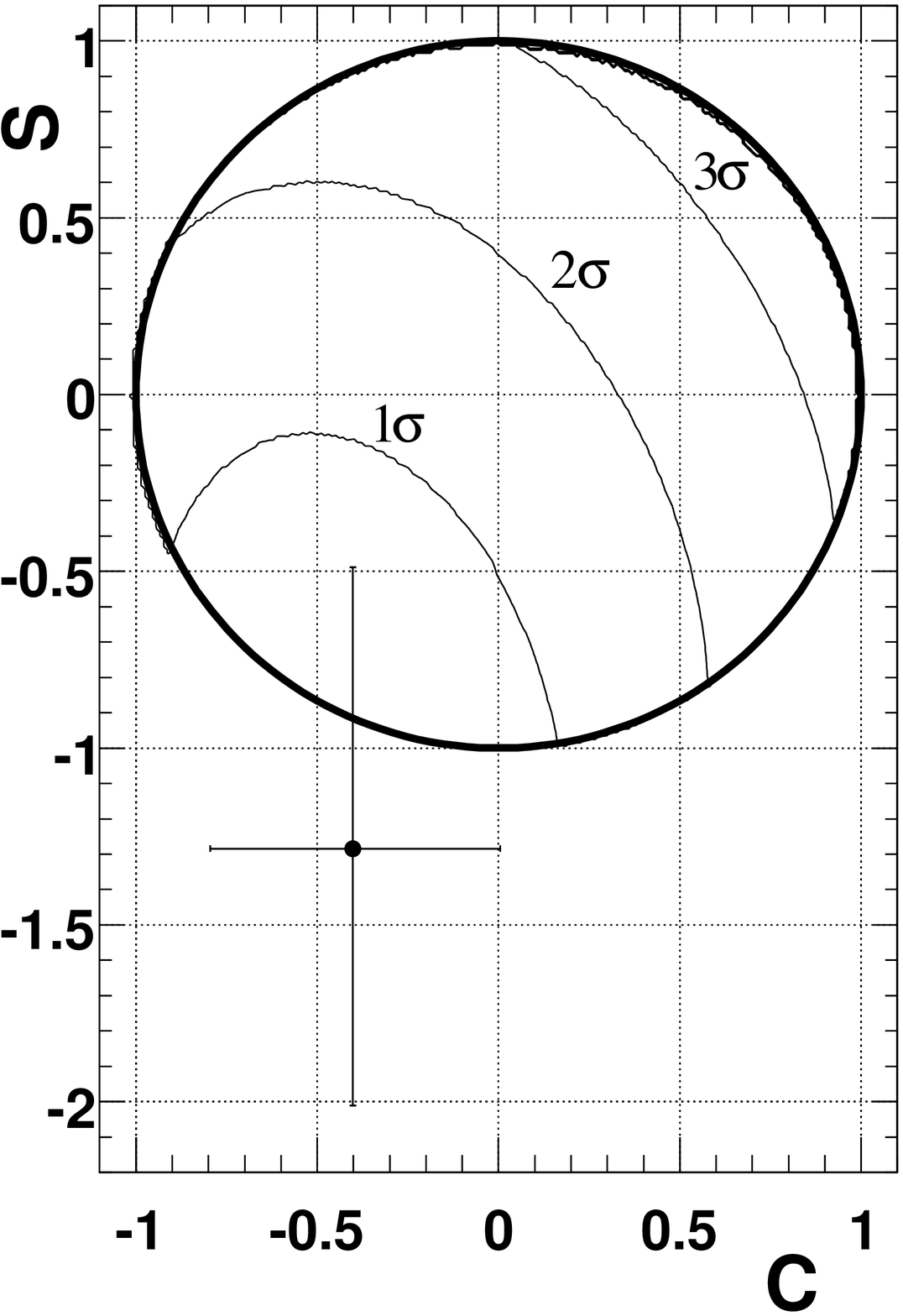}
\caption{Left: distributions of $\deltat$ for $\Bz\to\KS\KS$ decays in data 
tagged as $\Bz$ (top) or $\Bzb$ (middle), and the asymmetry (bottom).  The data is enhanced in signal decays using requirements on probability ratios.  The solid curve represents the PDF projection for the sum of signal and background, while the dotted curve shows the contribution from background only.  Right: 
Likelihood contours in the $S$ vs. $C$ plane, where
$n\sigma$ corresponds to a change in $-2\log{\cal L}$
of $2.3$ for $n=1$,
$6.2$ for $n=2$, and $11.8$ for $n=3$.  The circle indicates
the physically allowed region, while the point with error bars denotes the result of the fit to data.}
\label{fig:Asym}
\end{center}
\end{figure}

In Fig.~\ref{fig:sPlots} we compare data and PDFs using the event-weighting 
technique described in Ref.~\cite{sPlots}.  We perform fits excluding the 
variable being shown; the covariance matrix and remaining PDFs are used to 
determine a weight that each event is either signal (main plot) or background (inset).  The resulting distributions (points with errors) are normalized to the 
appropriate yield and can be directly compared with the PDFs (solid curves) 
used in the fits.  We find good agreement between data and the assumed 
shapes in both $\mes$ and $\de$.  In Fig.~\ref{fig:Asym} we display the $\deltat$ distributions for $\KS\KS$ events 
tagged as $\Bz$ or $\Bzb$, and the asymmetry 
${\cal A} = \left(N_{\Bz} - N_{\Bzb}\right)/\left(N_{\Bz} + N_{\Bzb}\right)$. The projections are enhanced in signal decays by selecting on probability ratios calculated from the signal and background PDFs (excluding $\deltat$).
The likelihood function in the $\Bz\to\KS\KS$ fit is used to derive Bayesian
confidence-level contours in the $C$ vs. $S$ plane by fixing $(S,C)$ to specific values, refitting the data, and recording the change in 
$-2\log{\cal L}$.  Figure~\ref{fig:Asym} shows the resulting $n\sigma$ 
contours in the physical region defined by $S^2+C^2 < 1$.

Systematic uncertainties on the signal yields are primarily due to the imperfect
knowledge of the PDF shapes.  We evaluate this uncertainty by varying the PDF 
parameters that are fixed in the fit within their statistical errors, and by 
substituting different functional forms for the PDF shapes.  For the charged 
modes, the largest contribution is due to the signal parameterization of 
$\mes$ and $\de$ ($3\%$ for $\KS\pip$, $4\%$ for $\KS\Kp$), while for the neutral mode it is due to the potential fit bias ($8.6\%$) determined from the pseudo-experiments.  We use the larger
of the value or uncertainty on the background asymmetries to set the
systematic uncertainty on ${\cal A_{\CP}}$ due to potential charge 
bias~\cite{KpiBaBar}.  We measure background asymmetries ${\cal A}_{\CP}(\KS\pip)=-0.010\pm 0.008$ and 
${\cal A}_{\CP}(\KS \Kp)=-0.005\pm 0.009$, which are consistent with no
bias and lead to a systematic uncertainty of $0.010$.  The dominant sources
of systematic uncertainty on $S$ and $C$ are due to
the positions of the means in $\mes$ and $\de$.  The statistical uncertainties of
the measured values of the \CP\ parameters are in good agreement with the 
expected error values ($0.8\pm 0.3$ for $S$ and $0.6\pm 0.2$ for 
$C$), while Monte Carlo studies confirm that the fit technique is unbiased 
for large values of the \CP\ parameters.

In summary, we have observed the decays $\Bu\to\Kzb\Kp$ and
$\Bz\to\Kz\Kzb$ with significances of $5.3\sigma$ and $7.3\sigma$, 
respectively.  The observed branching fractions are consistent with recent theoretical estimates~\cite{FleischerRecksiegel,BenekeNeubertAndYYKeumAndFleischerRecksiegel2}. The measured values of the time-dependent 
\CP-violating asymmetry parameters in the $\Bz\to\KS\KS$ mode reported here indicate
that large positive values of $S$ are disfavored, although more
data will be needed to confirm this result.  We have also improved our
measurements of the branching fraction and \CP-violating charge asymmetry 
in $\Bu\to\KS\pip$; both are consistent with previous measurements 
by other experiments~\cite{BelleAndCleoACP}.

We are grateful for the excellent luminosity and machine conditions
provided by our \pep2\ colleagues, 
and for the substantial dedicated effort from
the computing organizations that support \babar.
The collaborating institutions wish to thank 
SLAC for its support and kind hospitality. 
This work is supported by
DOE
and NSF (USA),
NSERC (Canada),
IHEP (China),
CEA and
CNRS-IN2P3
(France),
BMBF and DFG
(Germany),
INFN (Italy),
FOM (The Netherlands),
NFR (Norway),
MIST (Russia), and
PPARC (United Kingdom). 
Individuals have received support from CONACyT (Mexico), A.~P.~Sloan Foundation, 
Research Corporation,
and Alexander von Humboldt Foundation.

\end{document}

%% file: pubboard/authors_jul2006.tex
%
\author{B.~Aubert}
\author{M.~Bona}
\author{D.~Boutigny}
\author{F.~Couderc}
\author{Y.~Karyotakis}
\author{J.~P.~Lees}
\author{V.~Poireau}
\author{V.~Tisserand}
\author{A.~Zghiche}
\affiliation{Laboratoire de Physique des Particules, IN2P3/CNRS et Universit\'e de Savoie,
 F-74941 Annecy-Le-Vieux, France }
\author{E.~Grauges}
\affiliation{Universitat de Barcelona, Facultat de Fisica, Departament ECM, E-08028 Barcelona, Spain }
% ...
\collaboration{The \babar\ Collaboration}
\noaffiliation

%% file: paper.bbl
\begin{thebibliography}{99}
\bibitem{KsKl}
The decay $\Bz\to\Kz\Kzb$ proceeds in an $S$-wave, which
produces equal fractions of $\KS\KS$ and $\KL\KL$, but no 
$\KS\KL$, neglecting \CP\ violation in the kaon system.
\end{thebibliography}
